# Plasmon-Soliton

**Eyal Feigenbaum and Meir Orenstein**

Department of Electrical Engineering, Technion, Haifa 32000, Israel

meiro@ee.technion.ac.il

**Abstract**: Formation of a novel hybrid-vector spatial plasmon-soliton in a Kerr slab embedded in-between metal plates is predicted and analyzed with a modified NLSE, encompassing hybrid vector field characteristics. Assisted by the transverse plasmonic effect, the self trapping dimension of the plasmon-soliton was substantially compressed (compared to the dielectrically cladded slab case) when reducing the slab width. The practical limitation of the Plasmon-soliton size reduction is determined by available nonlinear materials and metal loss. For the extreme reported values of nonlinear index change, we predict soliton with a cross section of 300nm×30nm (average dimension of 100nm).

The downscaling of conventional photonics is halted when approaching a transverse dimension of ~$\lambda/2$ ($\lambda$ is the wavelength). This limitation can be alleviated by the incorporation of metals, giving rise to surface plasmon polaritons (SPP) [1,2]. Here we analyze a novel configuration where light is transversely tightly guided between two metal layers, and laterally self trapped by a nonlinear Kerr effect to yield a plasmon-soliton. The tight field confinement of this scheme is important for potential excitation of plasmon-solitons by a low power continuous wave optical source, which is further assisted by the expected small group velocity of the plasmon-soliton. In this letter we present for the first time both an analysis of TM spatial solitons, using the Non-linear Schrödinger Equation (NLSE), where the full vectorial field is considered, as well as the prediction and characteristics of SPP based solitons.

Defying the regular diffraction limits by SPP waves stems from their slow wave characteristics, namely a propagation momentum larger than that of a plane wave in the same dielectric. The specific setting of a dielectric layer enclosed by two metal layers, supporting an SPP mode, is thus equivalent to a 2 dimensional subspace with reduced in-plan diffraction. Both linear [1] and nonlinear photonics (this letter) with lateral light field size smaller than the wavelength are facilitated. From a modeling perspective - the reduced diffraction is instrumental for a proper transformation of Maxwell equations to the NLSE of the envelope function for ultra small solitons.

The plasmon-soliton analysis is preceded by a general derivation of a novel NLSE for transverse TM modes with lateral nonlinear confinement, taking into account the two electric field components, while the plasmonic effect is introduced later via a metal cladding structure. An NLSE analysis of TM solitons, laterally (in-plane) self trapped by the Kerr nonlinearity, while transversely (vertical) confined by a slab waveguide structure, was never performed, with comprehensive consideration of the vectorial field (to the best of our knowledge). This type of analysis is essential for SPP based solitons which carry a substantial longitudal electric field. Experimental reports of transversely guided solitons in slab waveguides examined only TE polarization [e.g. 3]. Theoretical analysis was focused on (1+1)D solitons with nonlinearity modifying the transverse guided mode profile (or self sustained in a bulk nonlinear media), while invariant (infinite) in the lateral dimension [e.g. 4-8]. In ref. [9] transverse dimension waveguiding effects were incorporated, using a variation method, however the full vector nature of the guided soliton was not included

In our structure - continuous waves propagating in z direction are confined in the x direction (transverse) by slab waveguide layers and self-trapped by Kerr nonlinearity in y axis (lateral). For vanishing nonlinearity the structure supports TE and TM modes, however, this distinction, based on $\partial y=0$, is false upon the on-set of the self-trapping which results in TE-TM mixing. Nevertheless, TE/TM based analysis is yet reasonable when the soliton width ($\Delta y$) is significantly larger than the mode width ($\Delta x$).

The wave equation obtained from Maxwell equations with Kerr nonlinearity [8] is:

$$\nabla \times \nabla \times \vec{E} - k^2 \vec{E} = \tfrac{4}{3} k_0^2 n_0 n_2 \left[ \left|\vec{E}\right|^2 \vec{E} + \tfrac{1}{2}\left(\vec{E} \cdot \vec{E}\right) \vec{E}^* \right] \qquad (1)$$

Where $k=k_0 n_0$ is material k-vector with $n_0$ and $n_2$ the respective linear and nonlinear Kerr refractive indices. For TM modes $\partial_z E_z = \partial_x E_x$, thus in (1) $\nabla \times \nabla \times \vec{E} = -\nabla^2 E_x \hat{x} - \nabla^2 E_z \hat{z}$, where the y-component is trivially satisfied. This implies that the wave equations for the two electric field components become identical, which is instrumental for the integrity of the soliton (otherwise each field component will evolve differently), as well as for the ability to employ a single scalar NLSE aligned with the field. The nonlinear index change is assumed to be smaller than the (very large) build-in interface index contrast, thus a negligible impact is expected on the plasmonic modal shape: $\vec{E} = \vec{E}_0(x) A(z,y) \exp\{j\beta z\}$. $\vec{E}_0$ is the TM field distribution for vanishing nonlinearity ($\vec{E}_0 = E_x \hat{x} + E_z \hat{z}$), $A$ a slowly varying envelope, and $\beta$ the $z$ propagation constant. By employing a paraxial approximation for $z$ propagation (namely omitting $A_{zz}$), Eq. (1) becomes:

$$2j\beta \vec{E}_0 A_z + \left[(k^2 - \beta^2)\vec{E}_0 + \vec{E}_0''\right] A + \vec{E}_0 A_{yy} + \tfrac{4}{3} k_0^2 n_0 n_2 \left[|\vec{E}_0|^2 \vec{E}_0 + \tfrac{1}{2}(\vec{E}_0 \cdot \vec{E}_0) \vec{E}_0^*\right] |A|^2 A = 0 \quad (2)$$

The constituents of the paraxial envelop are plasmonic modes, each carrying transverse and longitudal field components, forming a lateral paraxial beam with vector field characteristics. It should be emphasized that the longitudal components, usually absent in a paraxial approximation, are not a signature of a failed approximation, but are rather due to the basic plasmonic eigen functions. The lateral diffraction does not contribute to the longitudinal fields. Multiplying (2) by $\vec{E}_0^*$ and averaging over x yields a scalar wave equation for the amplitude $A$:

$$j2\beta A_z + \left[(k^2 - \beta^2) + I_2/I\right] A + A_{yy} + \tfrac{4}{3} k_0^2 n_0 n_2 (I_3/I) |A|^2 A = 0 \quad (3)$$

$$I = \int_{-\infty}^{+\infty} |\vec{E}_0|^2 dx; \quad I_2 = \int_{-\infty}^{+\infty} \vec{E}_0^* \vec{E}_0'' dx; \quad I_3 = \int_{core} \left\{|\vec{E}_0|^4 + \tfrac{1}{2}(\vec{E}_0 \cdot \vec{E}_0)(\vec{E}_0^* \cdot \vec{E}_0^*)\right\} dx$$

The averaging operation is reasonable when the transverse (x) cross-section is smaller than the lateral (y). For a slowly varying amplitude ($a = A\exp\{-j(k^2-\beta^2+I_2/I)/(2\beta)z)\}$) the NLSE is:

$$j a_z + (1/2\beta) a_{yy} + (2/3\beta) k_0^2 n_0 n_2 (I_3/I) |a|^2 a = 0 \quad (4)$$

A first order soliton is a solution of this equation, with peak amplitude η and width Δy:

$$\Delta y \cdot \eta = \sqrt{3I/(16\pi^2 n_0 n_2 I_3)}\, \lambda_0 \quad (5)$$

$\lambda_0$ is free space wavelength. We normalize the average vertical intensity in the core to unity ($d^{-1} \int_{core} |E_0|^2 dx = 1$), thus $\eta^2$ is the soliton average power.

For completeness we note that a similar derivation for TE modes indicates the deficiency of diffraction in the y direction. For TE mode ($\vec{E}_0 = E_y \hat{y}$) the $\nabla \times \nabla \times E$ in (1) equals $\partial_{xy} E_y \hat{x} - (\partial_{xx} + \partial_{zz}) E_y \hat{y} + \partial_{zy} E_y \hat{z}$, which for non cross-directional susceptibilities yields $\partial_y E_y=0$ - a manifestation to the lack of a soliton solution - self trapped in the y direction and supported by a TE mode in the x layered structure.

We employ the general results of TM based solitons to the gap SPP configuration, where a thin Kerr dielectric slab is sandwiched between two metal layers. The linear structure supports modes with transverse dimension much smaller than a wavelength. The modal field is effectively confined within the gap between the metal layers, and can be reduced almost indefinitely without cutoff by reducing the gap thickness. In Fig. 1(a) the effective index dependence on both the wavelength and gap size (d) is depicted. Symmetric modes are exhibiting higher effective indices (inferring to potentially smaller lateral solitons), for narrower gaps or for reduced wavelength.

By applying (5) for a non-plasmonic configuration, namely a regular dielectrically cladded nonlinear slab, the effective dimension of the spatial soliton ($D_{eff}=(\Delta x \, \Delta y)^{0.5}$) exhibits a minimum at a slab thickness of ~ $\lambda/2$ (see Fig. 1(b)). The soliton confinement can not be enhanced further by increasing the intensity, as the nonlinear index change $\Delta n$ reaches its saturation value [10-11]. However, for the metallic cladded nonlinear slab – as verified below - the "plasmonic" effect is shown to overcome this minimum effective soliton width restriction.

Solving for a nonlinear Kerr medium sandwiched between two silver layers yields a closed form expression for the soliton width. The effective slab mode size embedded in metal cladding is depicted in Fig. 1(b) (blue) and compared to that of air cladded slab (red). The Kerr media used in the calculations include nonlinear glasses and polymers [12-13], exhibiting a nonlinear index change of $10^{-3}$ as well as materials exhibiting an extremely high index change of 0.1 [10-11]. The "plasmonic" signature is evident: in contrast to the minimum of $D_{eff}$ exhibited for dielectric clad, here the effective size may be reduced with gap thickness and nano-scale dimensions below the diffraction limit ($D_{eff}<100nm$) are achievable. As a precaution – we restrict the validity of our model to regions where $\Delta y \gg \Delta x$ (not satisfied in the dashed segments of the figure).

The unique role of the diffraction in the plasmonic case is quantified by the factor $I_3/I$ which weights the nonlinear coefficient $n_2$ in Eq. 4. $I_3$ is the vertically integrated **squared** intensity in the core and I is the total integrated intensity. For a soliton in an all-dielectric slab, the mode confinement deteriorates rapidly as the slab thickness is reduced below the diffraction limit, and the decreased factor $I_3/I$ results in a wider soliton. However, in a plasmonic gap, the mode confinement is nearly thickness independent and the buildup of the intensity for reduced gap size contributes to an enhanced $I_3/I$ factor and to a narrower soliton. It is important to note that unlike the well known indefinite enhancement of the linear effective index with gap size reduction, the $I_3/I$ factor, as well the enhancement of the non linear index change are saturated (at a gap size of ~10nm for our parameters).

The cross-section of a "micrometer-size" plasmon soliton field, obtained for a nonlinear glass or polymer exhibiting a nonlinear index change of 0.005 are depicted in Fig. 2(a). Similar "long range" Plasmon-solitons can be obtained on a single thin metal layer embedded in the Kerr medium and even on a single interface between metal and a Kerr dielectric. The cross sectional dimensions are $\Delta x=1\mu m$, $\Delta y=6\mu m$, however to enable propagation distances of many millimeters – a loss of ~100cm$^{-1}$ should be compensated – e.g. by embedding gain in the medium.

To asses correctly the subsequent plasmon-soliton solutions we should note that: First – the typical propagation length for highly confined SPP waves is limited to the micro scale regime due to metal losses, and it is this typical propagation length we expect for the plasmon-soliton as well; Second, we discuss here only solutions that are within the constrains of the analytical model – namely formal hybrid NLSE plasmon-solitons, however other and not less-interesting nonlinear wave-packets may exist under the conditions set by the plasmonic effect – which require more complex modeling (e.g. for higher intensity, lower wavelengths, similar transverse and lateral dimensions, rapidly varying envelop etc.).

The two major physical sources limiting the compression of the Plasmon-soliton beam size are the maximal attainable nonlinear index change ($\Delta n$) and the mode attenuation. As the effective wavelength in the plasmonic gap is reduced, (lower free-space wavelength or narrower gaps), the mode attenuation becomes larger, setting a practical limit on the plasmonic assisted reduced diffraction. By compensating the mode loss [14], nano-scale effective beam size ($D_{eff}$) is expected even for $\Delta n$ values smaller than the extreme reported values, e.g. $D_{eff}$=100nm is achieved for $\lambda_0$=820nm d=12nm and $\Delta n$=0.01. Yet, the fundamental saturation of the factor $I_3/I$ for reduced gap sizes, limits the soliton lateral width ($\Delta y$) to the diffraction limit dimension. Therefore, plasmon soliton beams attainable may have lateral size of about the diffraction limit and transverse size of few 10s nm.

The limiting factors of the plasmon soliton size are put forward by the incorporation of media exhibiting the highest reported nonlinear index change $\Delta n$=0.1 [10-11] and a gap size of 30nm (Fig. 2(b)). Such a plasmon-soliton, excited by a wavelength of 820nm has nano-scale effective width of ~100nm, while an all-dielectric slab-soliton at the same power has a minimal effective width exceeding the diffraction limit ($D_{eff}$=470nm). Furthermore – the all dielectric soliton accumulates upon propagation nonlinear phase comparable to the linear phase of the optical carrier, which severely limits the soliton notion. For the plasmon-soliton the metal loss [15] results in a decay length of ~ 10μm, larger than the soliton length of ~5μm, which rationalizes the use of 'soliton'. Furthermore, checking the validity of our approximations – reaffirms both the paraxial approximation: the soliton anzats is solving also the nonparaxial equation, as the soliton length is much larger than the wavelength of the carrier wave (440nm), as well as the TM assumption - $\Delta y$~300nm is order of magnitude larger than $\Delta x$=30nm. Applying the same geometry and nonlinear index change, for the communications wavelength (1550nm) is shown in Fig. 2(c) to exhibit also a sub-half-wavelength effective width (~280nm) but with reduced losses - decay length of 30μm.

We derived a hybrid vector NLSE describing slab confined TM modes which are self trapped in the free (in-plan) dimension by the Kerr effect. A nonlinear slab sandwiched between metal layers supports a hybrid "Plasmon-Soliton", and exhibits high confinement with lower effective beam size than the conventional diffraction limit - although practically the soliton self-trapping dimension is not reduced below the diffraction limit with available non linear materials. A novel dispersion characteristic – the non linear index dispersion, exhibits saturation at small gap width – while the linear effective index keeps increasing indefinitely.

We would like to acknowledge the Israel Ministry of Science and Technology for a partial support of this research. We would like to thank Prof. M. Segev for helpful discussions.

**Figure Captions**

FIG. 1. (a) $n_{eff}$ of plasmonic gap vs. wavelength: even (bold) and odd (dashed) modes, single surface mode (dot-dashed). (b) $D_{eff}$ vs. thickness (d) of the nonlinear Kerr slab embedded in metal (blue) and air (red). $\lambda_0$=820nm, nonlinear index change $\Delta n$ is marked on curves. Diffraction limit is denoted by dashed (green) line. $\lambda_{plasma}$=137nm, $n_0$=1.55.

FIG. 2. Intensity distributions of Plasmon-soliton in Kerr medium embedded within silver layers [15]: (a) $\Delta n$=0.005, $\lambda_0$=1550nm, d=1μm, $\varepsilon_M$=-103.5-i10, $D_{eff}$=2.4μm. (b) $\Delta n$=0.1, $\lambda_0$=820nm, d=30nm, $\varepsilon_M$=-30.2-i1.6, $D_{eff}$=95nm. (c) $\Delta n$=0.1, $\lambda_0$=1550nm, d=100nm, $\varepsilon_M$=-103.5-i10, $D_{eff}$=283nm. $n_0$=1.5.

**Figure 1**

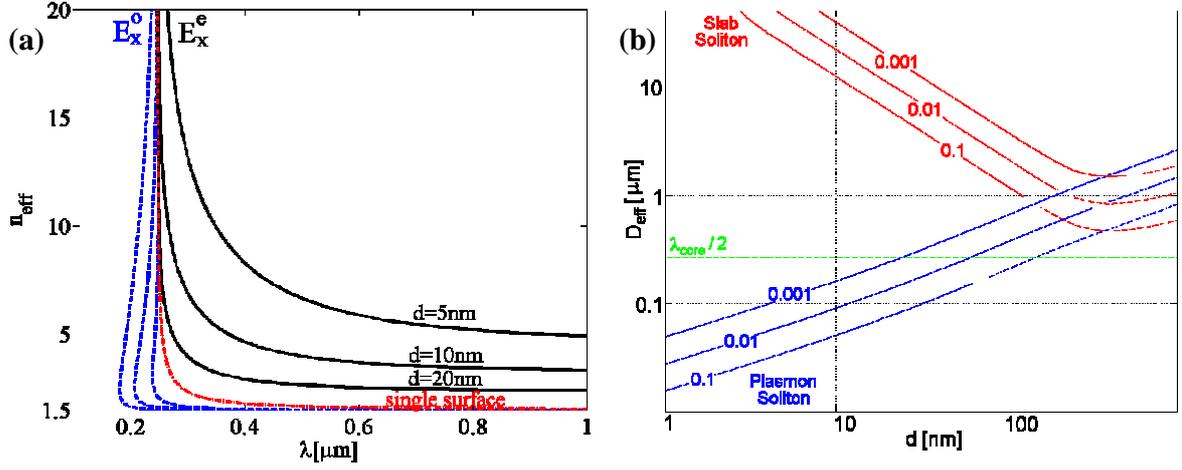

FIG. 1. (a) $n_{eff}$ of plasmonic gap vs. wavelength: even (bold) and odd (dashed) modes, single surface mode (dot-dashed). (b) $D_{eff}$ vs. thickness (d) of the nonlinear Kerr slab embedded in metal (blue) and air (red). $\lambda_0$=820nm, nonlinear index change $\Delta n$ is marked on curves. Diffraction limit is denoted by dashed (green) line. $\lambda_{plasma}$=137nm, $n_0$=1.55.

**Figure 2**

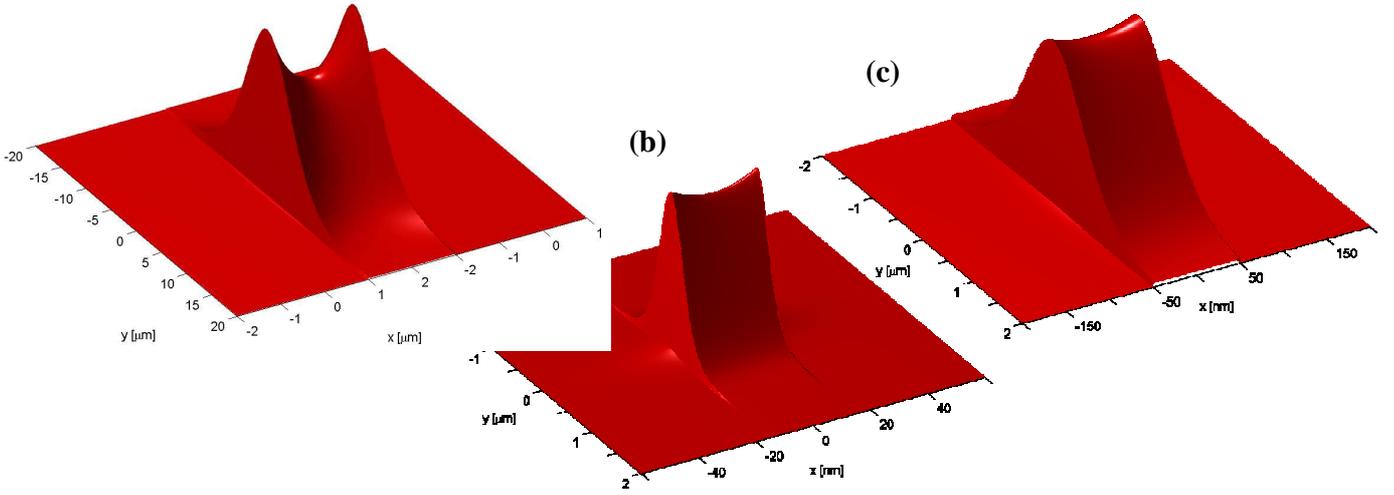

FIG. 2. Intensity distributions of Plasmon-soliton in Kerr medium embedded within silver layers [15]: (a) $\Delta n=0.005$, $\lambda_0=1550$nm, d=1µm, $\varepsilon_M=-103.5-i10$, $D_{eff}=2.4$µm. (b) $\Delta n=0.1$, $\lambda_0=820$nm, d=30nm, $\varepsilon_M=-30.2-i1.6$, $D_{eff}=95$nm. (c) $\Delta n=0.1$, $\lambda_0=1550$nm, d=100nm, $\varepsilon_M=-103.5-i10$, $D_{eff}=283$nm. $n_0=1.5$.